\documentclass[runningheads]{llncs}

\usepackage[T1]{fontenc}

\usepackage{graphicx}

\usepackage{ amsmath, amssymb, amsfonts, bussproofs, cmll, multicol,
  stmaryrd, hyperref, tabularx, listings, subcaption,
  listings, xcolor, bbold, mdframed }

\renewcommand{\vec}[1]{\overrightarrow{#1}}

\newcommand{\lambdaamor}{\texorpdfstring{$\lambda_{\text{amor}}$}{lambda-amor}}
\newcommand{{\dlpcf}}{\texorpdfstring{d$\ell$PCF}{dlPCF}}
\newcommand{\seq}{\vdash}

\newcommand{\lol}{\multimap}
\renewcommand{\phi}{\varphi}
\newcommand{\bdot}{\,\boldsymbol\cdotp\,}
\newcommand{\anypol}{\pm}
\newcommand{\ST}{\text{ST}}
\newcommand{\cmd}[2]{\langle{} #1 \parallel{} #2 \rangle{} }
\newcommand{\bigcmd}[2]{\Bigl\langle{} &#1 &\Big\Vert{} &#2 &\Bigr\rangle{} }
\newcommand{\redr}{\to}
\newcommand{\gramsep}{\;|\;}
\newcommand{\transl}[1]{\llbracket{} #1 \rrbracket}

\newcommand{\thunk}{{\Uparrow}}
\newcommand{\closure}{{\Downarrow}}
\newcommand{\unInf}[1]{ \UnaryInfC{\( #1 \)} }
\newcommand{\binInf}[1]{ \BinaryInfC{\( #1 \)} }

\newcommand{\ax}[1]{\AxiomC{ \( #1\) }}
\newcommand{\rulename}[1]{\RightLabel{\scriptsize (#1)}}
\newcommand{\typrule}[2][0.3]{
  \begin{subfigure}[c]{ #1\textwidth }
    \small
    \centering
    #2 \DisplayProof{}
    \vspace{1em}
  \end{subfigure}
}
\newcommand{\diff}[2]{#2}

\DeclareMathOperator{\andt}{\&}
\DeclareMathOperator{\barseq}{|}

\newcolumntype{L}{>{$}l<{$}} 
\newcolumntype{R}{>{$}r<{$}} 
\newcolumntype{C}{>{$}c<{$}} 

\lstdefinelanguage{LAME}{
  morekeywords={data, comput, with, of, where, |},
  morecomment=[s]{(*}{*)},
}

\begin{document}

\title{A reusable machine-calculus for automated resource analyses\thanks{This work has been partially performed at the IRILL center for Free Software Research and Innovation in Paris, France. }}
\author{Hector Suzanne\orcidID{0000-0002-7761-8779} \and Emmanuel Chailloux\orcidID{0000-0002-2400-9523}}
\authorrunning{H. Suzanne and E. Chailloux}
\institute{Sorbonne Université -- CNRS, LIP6\\\email{\{hector.suzanne,emmanuel.chailloux\}@lip6.fr}}

\maketitle

\begin{abstract}
  We introduce \diff{a Call-By-Push-Value abstract machine for automated
    resource analysis}{ an automated resource analysis technique is introduced,
    targeting a Call-By-Push-Value abstract machine}, with memory prediction as
  a practical goal. The machine has a polymorphic and linear type system
  enhanced with a first-order logical fragment, which encodes both low-level
  operational semantics of resource manipulations and high-level synthesis of
  algorithmic complexity.

  Resource analysis must involve a diversity of static analysis, for
  escape, aliasing, algorithmic invariants, and more. Knowing this, we implement
  the \emph{Automated Amortized Resource Analysis} framework (AARA) from scratch
  in our generic system. In this setting, access to resources is a state-passing
  effect which produces a compile-time approximation of runtime resource
  usage.

  \diff{We implemented type inference, constraint generation, and elaboration of
    bounds for our calculus.}{We implemented type inference constraint
    generation for our calculus, accompanied with an elaboration of bounds for
    iterators on algebraic datatypes, for minimal ML-style programming languages
    with Call-by-Value and Call-By-Push-Value semantics.} The closed-formed
  bounds are derived as multivariate polynomials over the integers. This now
  serves as a base for the development of an experimental toolkit for automated
  memory analysis of functional languages.

\end{abstract}

\section{Introduction}

Typed functional programming offers some structural safety out-of-the-box, but
correctness of systems also depends on quantitative, material concerns: memory
consumption must remain within bounds, latency and energy cost must be low, etc.
This highlights the need for general-purpose resource analysis tools for typed
functional languages. But functional languages in the style of ML or Haskell
pose specific challenges for resource predictions. First, dynamic allocations is
an inherent problem, since the size of the allocated data cannot be fully
determined statically. Second, prevalent use of garbage collectors and reference
counting in those languages means de-allocation points are implicit. Lastly,
closures (and high-order programming in general) purposefully hide the amount of
resources and state of data they close over.

Amortized complexity~\cite{tarjan1985} has been used to extend functional type
systems with resource analyses, notably in Hoffmann's \emph{Automated Amortized
  Resource Analysis (AARA)}~\cite{hoffmann2011}. But extending both the
precision and generality of current methods puts a large burden on formalisms
and implementations. In this paper, we develop a three-step approach to improve
this situation:
\begin{enumerate}
  \item \diff{We introduce a typed abstract machine with Call-By-Push-Value
        semantics, called the \textbf{ILL}-calculus, a Call-By-Push-Value
        machine typed with high order \textbf{I}ntuitionnistic \textbf{L}inear
        \textbf{L}ogic.}{We extend a Call-by-Push-Value abstract machine typed
        with intuitionistic linear logic introduced by Curien et al. } Programs
        are decomposed as interaction between values and stacks, whose types can
        involve \emph{parameters} variables, which denotes quantities of
        resources and number of combinatorial patterns in data. These parameters
        are guarded by first-order constraints. \diff{}{Call-by-Push-Value semantics
        strictly partition types into data-types and computation-types, and use
        \emph{closures} and \emph{thunks} to mediate between them.}
  \item Then, we devise a Call-By-Push-Value effects, which reflects at
        type-level the logical requirements inputs and outputs states of
        programs: execution go well for all states ($\forall$) with enough
        resources, and returns some arbitrary state ($\exists$) with some
        allocated resources.
  \item Finally, our implementation extracts a global constraint on resource
        usage from the type of the rewritten program. Those final constraints
        are expressed in first-order \diff{intuitionistic}{} arithmetic, and can
        be exported or solved automatically \diff{}{using our heuristic
        mimicking AARA reasoning. Solvers restricted to intuitionistic logic can
        elaborate resource expression back into the program}.
\end{enumerate}

Users can annotate higher-order code with domain-specific constraints which
partially specify runtime behavior. This allows for verification of high-order
programs through annotations in the general case, which is, to our knowledge,
novel in implementations of AARA. Using our system, resources analyses
can be decomposed into independent phases: a program is compiled into the
machine according to its CBPV semantics, it is then automatically rewritten with
our effect and typechecked. Finally, the domain-specific constraints obtained by
type-checking are solved using arithmetic solvers. Note that this last step does
not involve the original programs or the semantics of the programming language,
as opposed to previous work.

\diff{Furthermore, producing}{Producing} resources bounds has many non-trivial
requirements: one must unravel memory aliasing, lifetime of allocations,
algorithmic complexity and invariants, etc. When all those analyses interact, it
becomes beneficial to use a formalism that puts them all on equal footings, as
opposed to one dedicated uniquely to resource analyses. In our experiments,
combinations of those analyses are easier implement and verify thanks to
factorization, effect system, and core inference procedure.

\subsubsection{Plan} Section~\ref{sec:context} is dedicated to the wider context
of amortized static analysis, the AARA method, and its recent formalizations in
linear, Call-By-Push-Value $\lambda$-calculus. This ends with a more detailed
summary of our technical setup implementing AARA. In section~\ref{sec:machine},
we introduce our generic abstract machine adopting those advances, and follow
with its type system in section~\ref{sec:types}. We then explain how to encode
resource analyses for high-level languages as a Call-By-Push-Value effect in our
model in section~\ref{sec:analysis}. We discuss our implementation and
automation of analyses for functional languages in
section~\ref{sec:application}, and describe perspectives for further research in
the conclusion section~\ref{sec:conclusion}.

\section{Context and \emph{State of the Art}}\label{sec:context}

We begin by fixing some important notions regarding \emph{amortized algorithm
  analysis}~\cite{tarjan1985}, reviewing the AARA framework, and presenting its
most recent instantiations. Amortized algorithm analysis is a method for
determining the complexity of some operations on a data structure. Instead of
merely accumulating costs, it represent programs and their resources together as
a closed system, and \diff{rephrases}{characterizes} cost as the minimum of
total resource allowing the execution to proceed. We then present with the
\emph{Automated Amortized Resource Analysis} framework and its recent advances
using \emph{Call-By-Push-Value}. Once those concepts are set up, we end the
section with our setup for recovering AARA in a generic Call-By-Push-Value
system.

\subsection{Amortized Analysis}

Amortized analysis is an ubiquitous tool for the analysis of algorithms over the
entire lifetime of some structure, introduced by Tarjan~\cite{tarjan1985} to
determine asymptotic costs, but it applies just as well to quantitative
prediction. Foreseeing the rest of the paper, we will represent programs by
abstract machines and follow the nomenclature of previous
works~\cite{curien2016}. States of the abstract machine are \emph{commands}
$c \in \mathbf{C}$, and are made up of a \emph{value} $V \in \mathbf{V}$ and an
\emph{continuation stack} $S \in \mathbf{S}$ with syntax $c = \cmd{V}{S}$.
Semantics are given by deterministic, small-step reduction which will be assumed
to terminate throughout the paper. The execution of a program is therefore a
finite sequence of commands $c = {(c_{i})}_{i\le n}$.

\subsubsection{Costs}

A \emph{cost metric} is a function
$m : \mathbf{C} \times \mathbf{C} \rightarrow \mathbb{Z}$ giving a cost for a
transition $c \to c'$. When $m(c,c') \ge 0$ we call the cost a \emph{debit}, and
when $m(c,c') < 0$ we call it a \emph{credit}. Those credits do not occur for
some costs models like time and energy, which cannot be recouped. Models with
credits, like memory or currency, require credits and follow an extra condition,
mimicking absence of bankruptcy: all intermediate costs $\sum_{j\le i}m_{j}$
must be positive. Figuratively, this means that memory cannot be freed before
having been allocated, that currency cannot be spent from an empty account, etc.

For a sequence of deterministic reductions $(c_{i})_{i\le n}$, we write
$m_{i} = m(c_{i},c_{i+1})$ the cost of a reduction step, and $m(c)$ the total
cost of the reduction sequence. This total cost is the maximum of costs that can
be reached at an intermediate state, that is
$m(c) = \max_{i\le n}\sum_{j\le i} m_{j}$.

\subsubsection{Potential}

This formalism for costs can be reformulated as a matter of \emph{transfer of
  resources}, an idea originally put forward by Tarjan~\cite{tarjan1985}. Assume
given a fixed, sufficiently high amount of resources $P$ to run a program (pick
any $P \ge m(c)$). Each intermediate command $c_{i}$ has a
positive amount of \emph{allocated} resources $q_{i} = \sum_{j\le i}m_{j}$, and
a positive amount of \emph{free} resources $p_{i} = P - q_{i}$. At the beginning
of execution, we have $p_{0} = P$ and $q_{0} = 0$, and as reduction progresses,
we have the two inductive relations $p_{i} = p_{i+1} + m_{i}$ and
$q_{i} + m_{i} = q_{i+1}$. Therefore, $p_{i} + q_{i} = p_{0} + q_{0} = P$ is an
invariant of execution: resources are neither created nor lost, but preserved.

\subsubsection{Predicting Amortized Costs}

The ``potential'' point of view frames the problem of cost analysis as one of
invariant search: given $V \in \mathbf{V}$, find some function
$f : \mathbf{S} \to \mathbb{N}$ such that $f(S) \ge m(\cmd{V}{S})$.
Specifically, for each type for environment, we define numerical invariants
called \emph{parameters} (size, length, height, etc.) which gives a function
$\phi : \mathbf{S} \to \mathbb{N}^{k}$ taking each environment to its numerical
invariant, and call $\phi$ a \emph{parameterization} of $S$. Many such
parameterizations exists for the same datum. For example, a tree might be
parameterized by its size, its depth and width, or more complex combinatorial
data. Then given a parameterization $\phi$ of the runtime data, the amortized
complexity of $V$ is a function $P_{V} : \mathbb{N}^{k} \to \mathbb{N}$ such
that $P_{V}(\phi(S)) \ge m(\cmd{V}{S})$.

\subsection{Automated Amortized Resource Analysis and \emph{RAML}}

Hoffmann's \emph{Automated Amortized Resource Analysis} (introduced
in~\cite{hoffmann2011}, see~\cite{hoffmann2022} for a retrospective) is a
type-theoretic framework for resource usage inference in functional languages.
\diff{}{We give here a short introduction to AARA for non-recuperable costs.
  Nevertheless, both AARA and our methods support them.}

In Hoffmann's work, costs are represented by pairs
$p = (p_{\text{max}},p_{\text{out}})$ with $p_{\text{max}}\ge p_{\text{out}}$,
which means ``evaluation has a maximal cost of $p_{\text{max}}$ of which
$p_{\text{out}}$ are still allocated at the end''. They are endowed with
sequencing written additively. Judgements $\Gamma \seq_{p'}^{p} e : T$ means that
if $p$ resources are free before evaluating $e$, then $p'$ are available
afterward.

Parameters in AARA are linear combinations of specialized parameters called
\emph{indices}, which directly represent the number of some pattern in a
structure; For example, the base indices of a list $l$ are the binomial
coefficients $\binom{\emph{len}(l)}{k}$ with constant $k$, which count the
number of non-contiguous sublists of $l$ with length $k$. The weights with which
indices are combined are subject to a linear-programming optimization to derive
bounds on $p_{\text{max}}$ and $p_{\text{out}}$. AARA doesn't use a linear type
system. Instead, source programs must be \emph{syntactically affine}: every
variable is used at most once, and explicitly duplicated, which splits its
weights.

We show the rule for pairs below, exhibits an important property of AARA typing:
it encodes operational semantics of the specific source language within its
typing rules. Below for example, the cost $k_{\text{pair}}$ is payed from
$p+k_{\text{pair}}$, then $e$ is evaluated, then $e'$, yielding the sequence of
potentials $(p+k_{\text{pair}}) \rightarrow p \rightarrow p' \rightarrow p''$.
{\small
\[
  \ax{\Gamma \seq^{p}_{p'} e : A}
  \ax{\Gamma \seq^{p'}_{p''} e' : B}
  \binInf{\Gamma \seq^{p+k_{\text{pair}}}_{p''} (e,e') : A \times B}
  \DisplayProof
\]}

Instances of AARA cover different complexity classes~\cite{kahn2020}, some
aspects of garbage collection~\cite{niu2018} for pure functional programming,
and some aspects of imperative programming with mutable
arrays~\cite{lichtman2017}. Hoffmann et al. have implemented AARA in
\emph{Resource-Aware ML}~\cite{hoffmann2017} (RAML), a type system for a
purely-functional subset of OCaml that infers memory bounds, and supports
\diff{}{reasoning the the number of nodes in} algebraic datatypes, iteration on
lists, deeply-nested pattern matching, and limited form of closures. On those
programs, RAML can infer costs for a class of algorithms of polynomial
complexity. The key point allowing RAML to precisely bound memory usage of OCaml
programs is its compile-time representation of heap pointers, allowing it to be
aware of memory aliasing. \diff{But support for high-order programming could be
  improved: when using a functional argument in an iteration, its cost
  specification can be instantiated each cycle, but cannot be changed during
  iteration. We refine this by using a formalism in which changes to functional
  arguments can be represented as alterations to their evaluation context.}{RAML
  support high-order programming, if high-order arguments do not change during
  successive calls to the high-order function that uses them. Our
  continuation-passing, defunctionalized system allows for such changes to be
  represented as modification to the argument's evaluation context, with can in
  theory be tracked using the same tools as data structures.}

\subsection{\dlpcf{} and \lambdaamor}

On the other end of the spectrum, type systems inspired from program logics can
prove complex properties, even for non-terminating programs. Dal~Lago \&
Gaboardi's \dlpcf~\cite{dallago2013,dallago2011} is a type system for the
$\lambda$-calculus with integers and fixpoints (PCF for short), with a
highly-parametric, linear, and dependent type system. It is relatively complete
up to a solver for arithmetic. This can encode, for example, the number of
execution steps of programs in the Krivine machine in the presence of fixpoints,
but finding closed forms of \diff{the}{this} number of steps is undecidable.
This highlights the impossibility of typing the costs of fixpoint in the general
case.

Originally, \dlpcf{} could only bound accumulative costs, but subsequent work by
Rajani, Gaboardi, Garg \& Hoffmann~\cite{rajani2021} have recovered amortization
within this setting. The resulting system, \lambdaamor, is a family of program
logics, parameterized by a first-order theory describing resources. Changing
this theory tunes the resulting system to be close to the syntax-directed,
inferable, and amortized costs of AARA, or the recursion logic of \dlpcf.
Resources and costs are represented using two primitive type constructors.
Computation incurring costs are typed $\mathbb{M}_{I}A$ (with $I$ the cost), and
implement a cost-accumulating monad. The type of values of type $A$ holding
potential is $[I]A$, and implements the dual potential-spending comonad. This
system \diff{is heavily inspired by}{uses} Levy's
\emph{Call-By-Push-Value}~\cite{levy2003} formalism for encoding effectful
$\lambda$-calculi, \diff{}{with two apparent limitation: first,} it uses two
different reductions: one for cost-free expression, and second that performs
resource interaction on normal forms of the previous one. \diff{}{Second, the
  finer semantics of Call-by-Push-Value are only used in \lambdaamor{} to
  analyse programs with coarser Call-by-Value semantics}.

\subsection{\diff{}{Abstract Interpretation for Resource Analysis}}

While the current work focuses on type-based resource analysis and extension to
finer program semantics, other resource analysis techniques have been created.
We note the \emph{CioaPP} system for resource analysis~\cite{lopez-garcia2018},
based on \emph{abstract interpretation}. This techniques approximates the
semantics of values using \emph{abstract domains}. For example, and integer
could be abstracted by an union of interval, to which the integer is guaranteed
to belong. Multiple languages are supported through compilation to \emph{Horn
  clauses} in the style of logic programming: programs are represented by
predicated $C(\vec{x})$ linking their source and semantics, defined by relation
to other predicates in clauses
$C'_{1}(\vec{x}) \wedge \dots \wedge C'_{2}(\vec{x}) \Rightarrow C(\vec{x})$, in
which the $\vec{x}$ are all quantified universally. Abstract domains for
$\vec{x}$ can then be directly built using the clauses, and a purpose-built
\emph{fixpoint operator} for each domain that abstract iteration. \emph{Cioa/PP}
can be used to derive polynomial, exponential and logarithmic complexities, and
verify that programs manipulate resources according to quantitative bounds over
all inputs or a restriced domain. Implementations exists for many monotone
resource metrics, such as time, energy and \emph{gas} (execution fees for smart
contracts on blockchains)~\cite{perez2020}. To our knowledge, recuperable
resources aren't supported.

\subsection{Our Technical Setup}

\diff{This decomposition of}{Decomposing} AARA into a Call-By-Push-Value effect
allows for a simplified embedding of languages and programming paradigms:
Call-By-Push-Value programs explicitly define their evaluation order and allow
for mixed-style evaluation. We exploit this in the next section. Furthermore,
this allows embedding AARA's index languages into a \diff{classical}{mainstream,
  general-purpose} type system (sequent-style System-F) and simplifies
formalization. We'll describe our type system and how to encode resource
analyses in sections \ref{sec:types} and \ref{sec:analysis}. As a consequence of
those two changes, the vast literature of typechecking and type inference then
becomes directly applicable, which we discuss in section \ref{sec:application}.

\section{The ILL-calculus}\label{sec:machine}

We now introduce the \emph{polarized \textbf{ILL}-calculus}, an abstract machine
calculus due to Curien, Fiore \&
Munch-Macca\-gnoni\cite{curien2016,munch-maccagnoni2017}, which we extend with
algebraic datatypes, fixpoints, and explicit sharing. The name is a nod to its
type-level semantics which are exactly \emph{polarized \textbf{I}ntuitionnistic
  \textbf{L}inear \textbf{L}ogic}. At runtime, it is exactly a
continuation-passing abstract machine for the \emph{Call-By-Push-Value}
$\lambda$-calculus. This technical setup allows for a state-passing effects, and
an encoding resource manipulations at type-level. \diff{}{The core of the
  machine is taken as-is from \cite{curien2016}, and we introduce the following
  additions for our purposes: explicit sharing of variables; polymorphism;
  fixpoints; and a notation for thunks and closures.}

\subsection{Generalities}

At the term level, the \textbf{ILL}-machine is an abstract machine, whose terms
are described in ``\textbf{Terms (linear)}'', figure \ref{fig:term-syntax}. The
first line defines \emph{commands} $c$, which are a pair $\cmd{V}{S}$ of a value
$V$ and a stack $S$. Both parts are tagged with a \emph{polarity}: $+$ for data
and $-$ for computations. When a value and a stack interact in a command, we
call $V$ \emph{left side} and $S$ \emph{right side}.

Below commands in the figure, values and stacks are defined in matching pairs of
same polarity. Some values are built inductively from constructors
$V = K(\vec{V})$ and interact with pattern matching stacks with many branches
$S = \mu( K(\vec{x}).c \gramsep \dots)$. Dually, some stack defined inductively
and terminated with a \emph{continuation variable}, giving
$S = K_{1}(\vec{V_{1}}) \bdot K_{2}(\vec{V_{2}}) \bdot \dots \bdot \alpha$, and
interact with pattern-matching values
$V = \mu ( (K(\vec{x}) \bdot \alpha).c \gramsep \dots)$. The continuation
variable $\alpha$ stands for a yet-unspecified stack, and value variables $x$
stand for yet-unspecified values. This implements \emph{continuation passing}:
instead of \emph{returning} a value, programs \emph{jump to the current
  continuation} by passing it to a stack.

\begin{figure}[p]
\textbf{Terms (linear)}
\begin{mdframed}
\centering
 \begin{tabular}{R@{\hspace{0.5em}}C@{\hspace{0.5em}}C@{\hspace{0.5em}}C@{\hspace{0.75em}}C}
 c ::=& \multicolumn{4}{L}{\cmd{V^+}{S^+}^+ \gramsep \cmd{V^-}{S^-}^- \;\;\;\emph{(cut)}}\\ \\
  & \emph{(var)} & \emph{(let-val)} & \emph{(data)} & \emph{(closure)} \\
   V^{+} ::=& x^{+} & \diagup & K(\vec{V}^{+}) & \closure(V^{-})  \\
  S^{+} ::=& \alpha^{+} & \mu x^+.c & \mu \left(\overrightarrow{K(\vec{x}^{+}).c}\right) & \mu \closure (x^{-}).c \\ \\
  & \emph{(var)} & \emph{(let-stk)} & \emph{(computation)} & \emph{(thunk)} \\
  V^{-} ::=& x^{-} & \mu \alpha^-.c & \mu \left(\overrightarrow{(K(\vec{x}^{+}) \bdot \alpha^{-}).c}\right)
        & \mu(\thunk \boldsymbol\bdot \alpha^{+}).c \\
  S^{-} ::=& \alpha^{-} & \diagup & K(\vec{V}^{+}) \bdot S^{-} &  \thunk \bdot S^{+} \\

  \end{tabular}
  \end{mdframed}

\textbf{Terms (non-linear)}
\begin{mdframed}
  \centering
 \begin{tabular}{R@{\hspace{1em}}C@{\hspace{1em}}C@{\hspace{0.5em}}C@{\hspace{0.75em}}C}
 c ::=& \multicolumn{4}{L}{\cmd{V^+}{\mu \textbf{del}.c} \;\gramsep\; \cmd{V^+}{\mu \textbf{dup}(x,y).c}\;\;\;\emph{(structure)}} \\ \\

   &\emph{(sharing)} & \emph{(fixpoint)} \\
                 V^+ ::=& \mu ({!} \bdot \alpha^{-}).c & \mu (\mathbf{fix} \bdot \alpha^-).\cmd{\textbf{self}}{S^+} \\
                 S^+ ::=& ! \bdot S^{-} & \textbf{fix} \bdot S^-

  \end{tabular}
  \end{mdframed}

  \textbf{Reductions}
\begin{mdframed}
  \centering
  \begin{tabular}{R@{\hspace{1em}}RRLLLL}
    \textit{(let-stack)}
    & \bigcmd{\mu^{+}\alpha.c}{S}
    &\;\;\redr\;\; c[S/\alpha] \\[1em]
    \textit{(let-value)}
    &\bigcmd{V}{\mu^{-}x.c}
    &\;\;\redr\;\; c[V/x] \\[1em]
    \textit{(weakening)}
    &\bigcmd{V}{\mu\textbf{del}.c}
    &\;\;\redr\;\; c \\[1em]
    \textit{(contraction)}
    &\bigcmd{V}{\mu\textbf{dup}(x,y).c}
    &\;\;\redr\;\; c[V/x,V/y]\\[1em]
    \textit{(closure)}
    & \bigcmd{\closure(V)}{\mu \closure(x).c}
    &\;\;\redr\;\; c[V/x] \\[1em]
    \textit{(thunk)}
    &\bigcmd{\mu (\thunk \bdot \alpha).c}{\thunk \bdot S}
    &\;\;\redr\;\; c[S/\alpha]  \\[1em]
    \textit{(datatypes)}
    &\bigcmd{ K_{j}(\vec{V}) }{ \mu\left( \overrightarrow{K_{i}(\vec{x_{i}}).c_{i}}\right)}
    &\;\;\redr\;\; c_{j}[\vec{V}/\vec{x_{j}}] \\[1em]
    \textit{(computations)}
    & \bigcmd{ \mu \left( \overrightarrow{(K_{i}(\vec{x_{i}}) \bdot \alpha_{i}).c_{i}}\right)}{
      K_{j}(\vec{V}) \bdot S }
    &\;\;\redr\;\; c_{j}[\vec{V}/\vec{x_{j}},S/\alpha_{j}] \\[1em]
    \textit{(sharing)}
    &\bigcmd{ \mu ({!} \bdot \alpha).c }{{!} \bdot S}
    &\;\;\redr\;\; c[S/\alpha]\\[1em]
    \textit{(fixpoint)}
    &\bigcmd{\mu(\textbf{fix}\bdot \alpha).\cmd{\textbf{self}}{S}}{\textbf{fix}\bdot S'} \\[1em]
    &\multicolumn{6}{R}{\;\;\redr\;\; \cmd{ \mu(\textbf{fix}\bdot \beta)\cmd{\textbf{self}}{S[\beta/\alpha]}}{S}[S'/\alpha]} \\
  \end{tabular}
\end{mdframed}
\caption{\textbf{ILL}-machine: term-level syntax}\label{fig:term-syntax}
\end{figure}

\subsection{Linear Fragment}

The \textbf{ILL} machine works with \emph{linear substitutions} unless stated
otherwise. In this subsection, the machine encodes linear computations, which
preserve resources held by values \emph{by definition}. We now describe each
pair of compatible values and stack which use linear substitution. For each
following \textbf{(bold label)}, please refer to the corresponding definition in
\textbf{Terms} and reduction in \textbf{Reduction}.

\subsubsection{\emph{(let-value)} and \emph{(let-stack)}}

The machine manipulates values and stacks with binders: the stack $\mu^{+}x.c$
captures the data on the other side, and jumps to $c$ with $x$ bound, which can
be understood as ``$\mathtt{let}\; x = \dots \;\mathtt{in}\; c$''. Dually, the
term $\mu^{-}\alpha.c$ captures the evaluation context on the other side of the
command in $\alpha$ and jumps to $c$. Those are the two reductions in
\textbf{Reduction}, figure~\ref{fig:term-syntax}.

\subsubsection{\emph{(data)}}

Algebraic type constructors are defined as in functional languages \emph{à la}
OCaml or Haskell. They have value-constructors $K$ to build values with the
familiar syntax $K(\vec{V})$. Data structures then are consumed by
\emph{pattern-matching} stacks: for example, a type with two constructors
$K_{1}(-)$, and $K_{2}(-,-)$ matches with a stack
$\mu( K_{1}(x_{1}).c_{1} \gramsep K_{2}(x_{2},y_{2}).c_{2})$. Those two reduce
together by branching and binding variables:
\[ \cmd{K_{1}(V_{1})}{\mu( K_{1}(x_{1}).c_{1} \gramsep K_{2}(x_{2},y_{2}).c_{2})} \redr c_{1}[V_{1}/x_{1}].\]

\subsubsection{\emph{(computation)}}

The same way datatypes has values build inductively from constructors
$K(\vec{V})$, computation types have \emph{stacks} defined inductively from
constructors $K(\vec{V})\bdot S$. Functions are the prototypal example: the
function $A \lol B$ has a constructor $\lambda(V) \bdot (S)$ which
carries an argument of type $A$ and a continuation stack consuming A $B$. Then,
the $\lambda$-calculus call $f(V)$ corresponds to the command
$\cmd{f}{\lambda(V) \bdot S}$, where $S$ is the outer context of the call
(hidden in $\lambda$-calculus). The body of $f$ is a pattern-matching value: we
write $f = \mu (\lambda(x) \bdot \alpha).c$, in which $x$ binds the argument and
$\alpha$ binds the continuation. They interact by reducing as:
\[ \cmd{\mu(\lambda(x)\bdot\alpha).c}{\lambda(V)\bdot S} \rightarrow c[V/x,S/\alpha]. \]

Types with many stack constructors implement computation with many different
calls, each call sharing the same environment (think, in OOP, of an object with
multiple methods, all sharing the same instance variables).

\subsubsection{\emph{(thunks $\thunk$)} and\emph{ (closures $\closure$)}}

Closures and thunks implement local control flow, by delaying calls to
computations and generation of data. Thunks $\mu(\thunk \bdot \alpha).c$ are
commands $c$ blocked from reducing, with free stack variable $\alpha$. When
evaluated together with a stack $\thunk \bdot S$, they bind $S$ to $\alpha$, and
jump to $c$. Formally, they reduces as
$\cmd{\mu(\thunk \bdot \alpha^{+}).c}{\thunk \bdot S^{+}} \to c[S^{+}/\alpha^{+}]$.
The commands $c[S/\alpha]$ immediately evaluates the thunk, and eventually its
return value will interact with $S$. Closures go the other way around, and delay
launching computation. This allows, for example, to store a function within a
data structure. They are the symmetric of thunks: a closed
computation $\closure (V^{-})$ is opened with a blocked context
$\mu \closure (x^{-}).c$, which captures $V^{-}$ as $x^{-}$ and launches $c$,
which sets up a new evaluation context for it.

\subsection{Call-By-Value Semantics}

The canonical encoding of linear call-by-value functions $A \lol B$ into
Call-By-Push-Value translates them as $\closure(A \lol \thunk B)$. At the
term-level, the linear function $\lambda x.e$ becomes a closure $\closure$ over
a function $\mu(\lambda(x)\bdot\alpha)$, which defines a thunk
$\mu(\thunk \bdot \beta)$, which evaluates $e$. We write $\transl{-}$ the
Call-By-Push-Value embedding of a call-by-value term or type. Putting it all
together, we have:
\[ \transl{A \lol B} = \closure \transl{A} \lol \thunk \transl{B}\]
\[ \transl{\lambda x.e} = \closure\; \mu( \lambda(x)\bdot\alpha ). \cmd{\mu(\thunk \bdot \beta).\cmd{\transl{e}}{\beta}}{\alpha} \]

The main point of interest of those semantics of CBV in CBPV is that they are
extendable with effects, which are implemented as systematic rewriting of thunks
and closures. Those rewritings can be sequenced to refine effects. In the last
sections, we combine an effect of type-level tracking of quantities and one for
state-passing to recover AARA.

\subsection{Non-Linear Fragment}

In order to encode non-linear programs, including recursion, and track them at
type-level, we introduce variations to closures that encode shared values and
recursive values.

\subsubsection{\emph{(sharing)}}

Shareable data is encoded as \emph{shareable commands} $\mu (! \bdot \alpha).c$,
whichare shared as-is then pass data to a stack $! \bdot S$. Linearity is
enforced at the type level by making them have a distinct type, and asking that
all value-variables bound in shared commands also have a shareable type. In
order to track non-linear substitutions, sharing is explicitly implemented as
stack matching shared values. The stack $\mu \mathbf{del}.c$ implements
weakening by silently ignoring the value it matches on, and
$\mu \mathbf{dup}(x,y).c$ implements contraction by binding two copies $x$ and
$y$ of the shared value.

\subsubsection{\emph{(fixpoint)}}

Finally, recursive computations are encoded as \emph{fixpoints}. They are also
subject to weakening and contraction, which enables the usual recursion schemes
of $\lambda$-calculus to be encoded into \textbf{ILL}. Formally, the stack
constructor $\mathbf{fix}\bdot S$ opens a fixpoint closure, expands its
definition in its body once, and feeds the resulting computation to $S$. On the
other side, fixpoints have syntax
$\mu(\mathbf{fix}\bdot\alpha).\cmd{\mathbf{self}}{S}$, where $\mathbf{self}$ is
a hole to be filled with the recursive value, and $S$ captures the self-referent
closure once filled-it and returns the defined recursive computation to
$\alpha$. The reduction substitutes the entire fixpoint into \textbf{self},
which copies $S$. This is made formal in the associated rule in
\textbf{Reduction}, figure~\ref{fig:term-syntax}. \diff{}{Note the
  $\alpha$-conversion in this rule, which protects one of the copies of $S$ from
  unwanted substitution.}

\section{Type System}\label{sec:types}

The end game of the type system is to derive a first-order constraint $C$ over
relevant quantities of a program, from which we then derive a bound. We call
those quantities \emph{parameters}. They represent amount of liquid resources,
or combinatorial information on data and computation. In this paper, we focus on
parameterizing data, for brevity.

It is capital that computations operate on data of arbitrary parameters. For
example, fixpoints will call themselves with arguments of varying sizes to
encode iteration. This means polymorphism over size \emph{must} be accounted for. We
solve this issue by bundling quantified parameters within constructors.

\subsection{Generalities}

At the type level, \textbf{ILL} is polarized \emph{intuitionistic linear sequent
  calculus}. Its syntax is described in \textbf{Types} and \textbf{Parameters},
figure 2. The types $A,B,C$ of values and stack can have two base sorts
$\mathcal{T} \in \{+,-\}$ reproducing their polarity. We also have parameters
$n,p,q$, with sorts in $\mathcal{P}$ which includes the integers $\mathbb{N}$.
Types can depend on parameters: they have sorts
$\vec{\mathcal{P}}\to \mathcal{T}$. Finally, type constructors are polymorphic
over types and parameters: they have sort
$\vec{\mathcal{T}} \to \vec{\mathcal{P}} \to \mathcal{T}$. For example, lists
can have heterogeneous parameters for each element: a list
$[a_{0};a_{1};\dots;a_{n}]$ which each $a_{i}$ of type $A(i)$ has type
$\mathtt{List}(A,n)$, with the arguments having type $A(n-1)$ for the head, then
$A(n-2)$, all the way to $A(0)$. The associated type constructor is
$\mathtt{List} : (\mathbb{N}\to +) \to \mathbb{N} \to +$.

\subsubsection{Primitives and Parameters}

The usual connectives of instuitionistic linear logic are definable as type
constructors: $\otimes$ and $\mathbb{1}$ (``pattern matching pairs'' and ``unit
type''), $\oplus$ and $\mathbb{0}$ (``sums'' and ``empty type''), $\andt$ and
$\top$ (``lazy pairs'' and ``top''), and $\lol$ (``linear functions'').
\diff{}{Thunks and closures are given their own types: closures over a
  computation $A^{-}$ have the positive type $\closure A^{-}$, and thunks
  returning some data typed $A^{+}$ have negative type $\thunk A^{+}$.} We also
take as given the integers $(\mathbb{N},0,1,+,\times)$ for parameterization.
Those can be extended to any first-order signature.

\subsubsection{Judgements}

Judgements are sequents $\Theta;C;\Gamma \seq \Delta$, which represent a typed
interface: inputs are denoted by value variables
$\Gamma = (\overrightarrow{x:A})$, and output by one stack variable
$\Delta = (\alpha : A)$. The parameters of this interface are
$\Theta = (\overrightarrow{p:\mathcal{P}})$ and are guarded by a first-order
constraint $C$. The parameters in $\Theta$ are bound in $C$, $\Gamma$ and
$\Delta$, and denote free quantities than be tuned within limits given by $C$.
Given this, the three judgements, for values $V$, stacks $S$ and commands $c$
are: \vspace{1em}
\begin{center}
\begin{tabular}[c]{|c|C|c|}
  \hline
  \emph{Syntax}& \emph{Sequent} & \emph{Given \boldmath$\Theta$ such that \boldmath$C$, we have \dots } \\
  \hline
  Values & \Theta; C; \Gamma \seq V : A &\emph{ a value \boldmath$V$ of type
                                                   $A$ in context $\Gamma$}. \\
 Stacks & \;\;\Theta; C; \Gamma \barseq S : A \seq \Delta \;\;&\emph{ a stack
                                                                  \boldmath$S$ of type $A$ in context $\Gamma,\Delta$}. \\
  Commands\;\; & c : (\Theta; C; \Gamma \seq \Delta) & \emph{a valid command \boldmath$c$
                                                            under context $\Gamma,\Delta$}.\;\; \\
  \hline
\end{tabular}
\vspace{1em}
\end{center}

The central rules of the type system are shown in ``\textbf{Example rules}''
figure~\ref{fig:types}. Commands are built in the (cut) rule by matching a value
and stack on their type and taking a conjunction of their constraints. Rules
($\mu L$) and ($\mu R$) are for binders. For example, ($\mu R$) turns a command
$c$ with a free variable $x:A$ into a stack $\mu x.c : A$ which interacts with
values of type $A$ by substituting them for $x$.

\begin{figure}[p]

  \textbf{Types}
  \begin{mdframed}
    \vspace{-1em}
  \begin{align*}
    \mathcal{T} &::= {+}  \gramsep {-} \gramsep \mathcal{P} \gramsep \mathcal{P} \rightarrow \mathcal{T} \\
    A &::= p \gramsep T_{\text{cons}}(\vec{A}) \gramsep \closure A \gramsep \thunk A \gramsep !A \gramsep \mathbf{Fix}\; A \\
    A &::= \mathbb{1} \gramsep A \otimes A \gramsep \mathbb{0} \gramsep A \oplus A \gramsep \top \gramsep A \andt A \gramsep A \lol A \gramsep \exists \Theta [C].C \gramsep \forall \Theta[C]. A \phantom{\vec{X}} \emph{(definable)}
  \end{align*}
\end{mdframed}

  \textbf{Parameters}
  \begin{mdframed}
    \vspace{-1em}
  \begin{align*}
    \mathcal{P} &::= \mathbb{N} \gramsep \dots \\
    p &::= 0 \gramsep 1 \gramsep p + q \gramsep \dots \\
    C &::= \mathbb{T} \gramsep \mathcal{R}(\overrightarrow{p})  \gramsep p = p
        \gramsep C \wedge C \gramsep C \Rightarrow C
        \gramsep \exists \overrightarrow{p}.\,C
        \gramsep \forall \overrightarrow{p}.\,C
  \end{align*}
  \end{mdframed}

\textbf{Example type definition: lists}
\begin{mdframed}
\begin{lstlisting}[mathescape=true,language=LAME]
data List(A:$\mathbb{N}\to {+}$, n:$\mathbb{N}$) =
  | Cons of A(m) $\otimes$ List(A,m) where $(m:\mathbb{N})$ with $n=m+1$
  | Nil with $n=0$  . (*no 'where' clause for Nil*)
\end{lstlisting}
\end{mdframed}

\textbf{Example type definition: state token}
\begin{mdframed}
\begin{small}
\begin{lstlisting}[mathescape=true,language=LAME]
data ST($p$,$q$ : $\mathbb{N}$) =
| init where $q = 0$
| debit$_{k}$   of ST($p'$,$q'$) with $p',q':\mathbb{N}\;\;\;\;$ where $p'+k=p \wedge q+k=q'$
| credit$_{k}$ of ST($p'$,$q'$) with $p',q':\mathbb{N}\;\;\;\;$ where $p+k=p' \wedge q'+k=q$
| slack   of ST($p'$,$q'$) with $p',q',k:\mathbb{N}$ where $p'+k=p \wedge q+k=q'$
\end{lstlisting}
\end{small}

\end{mdframed}
\textbf{Example rules: identity and lists}
\begin{mdframed}
  \centering
  \typrule[0.35]{
       \ax{c : (\Theta; C; \Gamma \seq \alpha:A^-)}
       \rulename{$\mu$L}
       \unInf{\Theta; C; \Gamma \seq \mu^- \alpha.c : A^-}
     }
     \typrule[0.35]{
       \ax{c : (\Theta; C; \Gamma, x:A^+ \seq \Delta)}
       \rulename{$\mu$R}
       \unInf{\Theta; C; \Gamma \barseq \mu^+ x.c : A^+ \seq \Delta}
     }

    \typrule[0.6]{
       \ax{\Theta; C; \Gamma \seq V : A^\anypol}
       \ax{\Theta'; C'; \Gamma' \barseq S : A^\anypol \seq \Delta}
       \rulename{cut}
       \binInf{\cmd{t}{e}^\anypol : (\Theta,\Theta'; C \wedge C'; \Gamma,\Gamma' \seq \Delta)}
     }

  \typrule[1]{
    \ax{\Theta_1; C_1; \Gamma_1 \seq V_1 : A(m)}
    \ax{\Theta_2;C_2; \Gamma_2 \seq V_2 : \texttt{List}(A,m)}
    \rulename{\texttt{Cons}R}
    \binInf{\Theta_1,\Theta_2; C; \Gamma_1, \Gamma_2 \seq \texttt{Cons}(V_1, V_2) : \texttt{List}(A,n)}
    \noLine
    \unInf{C = \exists m. (n=m+1) \wedge C_1 \wedge C_2}
  }

  \typrule[1]{
    \ax{c_1 : (\Theta;C_1;\Gamma \seq \Delta)}
    \ax{c_2 : \left(\Theta,m:\mathbb{N}; C_2; \Gamma, x:A(m), y:\texttt{List}(A,m) \seq \Delta\right)}
    \binInf{ \Theta;C;\Gamma \barseq \mu (\; \texttt{Nil}().c_1 \barseq \texttt{Cons}(x,y).c_2\;) : \texttt{List}(A,n) \seq \Delta }
    \noLine
    \unInf{C = ((n = 0) \Rightarrow C_1) \wedge (\forall m. (n=m+1) \Rightarrow C_2)  \;\;\;{\scriptsize(\texttt{List}\text{L})}}
  }
\end{mdframed}
\caption{\textbf{ILL-machine: type system and examples}}\label{fig:types}
\end{figure}

\subsection{Datatypes with Parameters}

Building up the logical constraint for resources and algorithmic invariants is
done by accumulating generic information about constructors of values and stack.
To present how this machinery works, we encode a very simple constraint:
\emph{a list is always one element longer than its tail}. The corresponding type
definition for lists, and resulting rules are shown in \textbf{Example type
  definition}, figure~\ref{fig:types}.
Lists have type $\mathtt{List}(A,n)$ with a type parameter $n:\mathbb{N}$
denoting size. The definition of the \texttt{Cons} constructor is reproduced
below.
\begin{lstlisting}[mathescape=true,language=LAME]
  | Cons of A(m)$\otimes\mathtt{List}$(A,m) where $(m:\mathbb{N})$ with $n=m+1$
\end{lstlisting}
The definition states that lists $\mathtt{List}(A,n)$ have a head of type $A(m)$ and a
tail of type $\mathtt{List}(A,m)$. \diff{}{The type of list elements is
  $A : \mathbb{N} \to +$, which allows each element to be given a distinct
parameterization according to its position. For example,
$\mathtt{List}(\mathtt{List}(\mathbb{N},{-}),n)$ is a type of lists of lists of
integer of decreasing lengths: the first list has size 10, the next one 9, etc.}
Formally, the parameter $m$ is introduced in the \textbf{where} clause, and is
guarded by the first-order constraint $n=m+1$ in the \textbf{with} clause. This
is to be understood as ``\textbf{where} $m$ is fresh integer parameter
\textbf{with} $n=m+1$''. When constructing a list $\mathtt{List}(A,n)$ with the
rule (\texttt{Cons}R), $m$ is added to $\Theta$ and the constraint $n=m+1$ is
added. Symmetrically, when pattern matching on a list $\mathtt{List}(A,n)$ with
the rule (\texttt{List}L), the branch of the pattern matching for \texttt{Cons}
must be well-typed for any $m$ such that $n=m+1$, which yields a constraint
$\forall m. (n=m+1) \Rightarrow C'$ in which $n=m+1$ is assumed.

\subsection{Implementing Polymorphism}

Once some parameterization of data is chosen, we want parameters-aware data and
computations. Recall that when introducing constructors, some parameters
$\Theta$ satisfying some constraint $C$ are introduced existentially, and when
this constructor is matched on, they are introduced universally. This allows us
to encode polymorphism over parameters as types with only one constructor. We
define existential quantification over parameters $\Theta$ such that $C$ with
the $\exists \Theta [C].A$ datatype as follows:
\begin{lstlisting}[mathescape=true,language=LAME]
data $\exists \Theta[C]$.A = $\mathtt{Pack}_{\Theta;C}$ of A($\Theta$) where $\Theta$ with $C$
\end{lstlisting}

Introducing $\mathtt{Pack}_{\Theta;C}(V)$ produces the
constraint $\exists \Theta. C$, and when
pattern-matching on it, $C$ is assumed to hold for some unknown $\Theta$, giving
a constraint $\forall \Theta. C \Rightarrow C''$. Universal quantification goes
the other way around, binding some constraint existentially in stacks, and
universally in values. Compile-time information $\Theta,C$ about a continuation
stack $S:A^{-}(\Theta)$ is witnessed by the stack constructor
$\mathtt{Spec}_{\Theta;C} \bdot S$. On the value
side, $\mu(\mathtt{Spec}_{\Theta;C} \bdot \alpha).c$ requires that $C$ in the
command $c$, and generates $\forall \Theta. C \Rightarrow C'$.

Closures over universally quantified computations take
any input such that some constraint holds. Likewise, thunks over existential
quantification type delayed computations with (yet undetermined) parameters. For
example, a thunk which returns a pair of lists whose total length is 10 can be
typed as
$ \thunk \exists(n,m:\mathbb{N})[n+m=10].\; \texttt{List}(A,n) \otimes \texttt{List}(A,m). $

\subsection{An Example of Encoding: \emph{append}}

A minimal, non-trivial example of parameter polymorphism is the \texttt{append}
function on lists. We implement it in two phases. First, we implement
\texttt{rev\_append}, which flips the first lists and appends it to the second.
Then, we define \texttt{append} with two calls to \texttt{rev\_append}. This
decomposition shows the function of parameter polymorphism, as both call to
\texttt{rev\_append} occur on lists of different sizes. Figure
\ref{fig:rev-append} lists the original ML code for both functions and the
compiled version of \texttt{rev\_append}. We omit giving the translation of
\texttt{append}, since it is straightforward once given \texttt{rev\_append}.

\texttt{rev\_append} is defined as a fixpoint. The first line in the definition
binds a self-reference to $f$ and binds a continuation $\alpha$ to which it
returns the function. In $c_{1}$, when the function is called for lists $l_{1}$
of size $n$ and $l_{2}$ of size $m$, the execution context built by the caller
instantiates the sizes with $\forall_{(n,m)}$ which the callee matches on. Then,
in $c_{2}$, we pattern-match on $l_{1}$, which introduces $\exists n'.n=n'+1$
(resp. $n=0$) if the list has a head (resp. is empty). In this last case, the
function recurses on itself in $c_{3}$. This recursive call is done with an
execution context $\forall_{(n',m+1)} \bdot S$, in which the new values
for $n$ and $m$ are instantiated.
\begin{figure}[t]
  \centering
  {\small

\begin{tabular}{L}
  \texttt{let rec rev\_append l1 l2 = match l1 with [] -> l2} \\
  \texttt{\hspace{2em}| h::t -> rev\_append t (h::l2)} \\
  \texttt{let append l1 l2 = rev\_append (rev\_append l1 []) l2}
\end{tabular}\hspace{3em}

\begin{align*}
  &\hspace{-2em}\texttt{rev\_append}
   :\; \mathbf{Fix}\; \forall n,m,A.\; \mathtt{L}(A,n) \lol \mathtt{L}(A,m)\lol \thunk \mathtt{L}(A,n+m) \\
   &= \mu( \mathbf{fix} \bdot \alpha). \cmd{\mathbf{self}}{\mu f.c_{1}} \\
  c_{1} &= \cmd{\mu(\forall_{(n,m)} \bdot l_{1} \bdot l_{2} \bdot \thunk \bdot \beta).c_{2}}{\alpha}\\
  c_{2} &= \cmd{l_{1}}{\mu(\; \mathtt{Nil}_{(n=0)}.\cmd{l_{2}}{\beta}
          \;|\; \mathtt{Cons}_{(\exists n'.n=n'+1)}(h,t).c_{3} \;)}\\
  c_{3} &= \cmd{f}{ \mathbf{fix} \bdot \forall_{(n',m+1)} \bdot t \bdot \mathtt{Cons}_{(m+1)}(h,l_{2}) \bdot \thunk \bdot \beta }
\end{align*}}
  \caption{\textbf{BILL} Source code of the \texttt{rev\_append} function }
  \label{fig:rev-append}
\end{figure}

\subsection{Soundness}

Reduction preserves parameterizations in the following sense:
\begin{theorem}
  If $c : (\Theta;C;\Gamma \seq \Delta)$ reduces as $c \to c'$, and
  $c' : (\Theta';C';\Gamma' \seq \Delta')$, then $\Theta' \subset \Theta$,
  $\Gamma' \subset \Gamma$, $\Delta = \Delta'$, and for every instantiation of
  $\Theta$, $C \Rightarrow C'$.
\end{theorem}

The proof is done by induction over typed reduction of commands, after proving
the standard Barendregt properties \diff{}{(which can be done following
  \cite{munch-maccagnoni2017})}. We only have space to briefly summarize the
salient point. First, we prove the statement for $\cmd{\mu \alpha.c}{S}$ and
$\cmd{V}{\mu x.c}$. Then, the only significant cases are
$\cmd{\mathtt{Pack}_{\Theta;C}(V)}{\mu \mathtt{Pack}_{\Theta;C}(x).c}$ and its
dual with \texttt{Spec}. The \texttt{Pack} command reduce to $c[V/x]$ and
generates the constraint
$(\exists \Theta. C) \wedge (\forall \Theta. C \Rightarrow C')$ with $C'$ the
constraint for $c$. This immediately implies $C \wedge C'$, which is also
constraint generated by $\cmd{V}{\mu x.c}$, and therefore $c[V/x]$. The case of
$\mathtt{Spec}$ is purely identical.

\section{Embedding AARA as an Effect}\label{sec:analysis}

In section 3, we refined the CBPV embedding of CBV functions to polymorphic
closures and thunks. This allowed to track parameters as control flow switches
in and out of programs. To recover AARA, we merely need to specialize this
translation to track sizes and resources.

\subsection{An Effect for Parameters}

With our setup, we can translate CBV programs to \textbf{ILL}-machine that
associates a constraint $C$ on their free parameters $\Theta$. This is
implemented by refining closures and thunks. Closures $\closure A$ are replaced
by \emph{closures over quantified computations}
$\closure \forall \Theta[C].A(\Theta)$, that is, computations that take
\emph{any} arguments with parameters $\Theta$ satisfying $C$. On the other side,
thunks $\thunk B$ are replaced with $\thunk \exists \Theta'[C'].B(\Theta')$,
which returns data parameterized by $\Theta'$ such that $C'$. With this effect,
the call-by-value linear function $A \lol B$ is translated to a parameter-aware
version that accept \emph{all} ($\forall$) inputs with the right parameters and
return \emph{some} ($\exists$) output with its own parameters. For example, the
function \texttt{append} on lists has a length-aware type (here in long
form, to show the implicit $\exists$ binder):
\[ \closure \forall n,m.\; \mathtt{List}(A,n) \lol \mathtt{List}(A,m) \lol \thunk \exists k[k=m+n].\; \mathtt{List}(A,k) \]

\subsection{Polymorphic State-Passing Effect}

We extend the translation of call-by-value functions with another effect: state
passing. Closures now accept a token $\ST(p,q)$ with $p$ free resources and $q$
allocated resources, and thunks return a token $\ST(p',q')$.
Closures $\closure \forall \Theta[C].(-) $ become
$\closure \forall \Theta[C]. \ST(p,q)\lol (-)$, in which $C$ guards the
resources $p$ and $q$. This means closures take in \emph{any} state whose
resources satisfy $C$. Likewise, thunks becomes
$\thunk \exists \Theta[C']. \ST(p',q') \otimes (-)$, and return resources $p'$
and $q'$ specified by $C'$.

This can lift the hidden inner behavior of \texttt{append} at type-level.
Relying on linear typing, the effects detect that the progressive deallocations
of the intermediate reversed list compensate for the progressive allocation of
the final result (under an ideal garbage collector). When typed with
state-passing below, the tokens' types shows that only $n$ nodes are
allocated \emph{simultaneously} for the call.
\begin{multline*}
\closure \forall n,m,p,q.\; \ST(p+n*k_{\mathtt{Cons}},q) \lol \mathtt{List}(A,n) \lol \mathtt{List}(A,m) \\ \lol \thunk \exists k[k=m+n].\; \ST(p,q+n*k_{\mathtt{Cons}}) \otimes \mathtt{List}(A,k)
\end{multline*}

\subsection{Token Encoding}

To implement this translation without specific primitives, we define a state
token capable of representing the operations of \emph{debit} (spending
resources), \emph{credit} (recovering resources), and \emph{slack} (aligning
costs upwards) at type-level. We define a type constructor $\ST$ with two
resource parameters, that implements those operations, in \textbf{Example type
  definition: state token}, figure~\ref{fig:types}.

The token begins its life as \texttt{init}, which has type
$\ST(p,0)$. Debiting $k_{0}$ resources from a token $s : \ST(p+k_{0},q)$
is done by using the constructor $\mathtt{debit}_{k_{0}}(s)$ which creates a new
token of type $ST(p,q+k_{0})$. The \texttt{credit} constructor implements the
opposite operation.
For slack, the amount of resources being wasted $k$ is left free. At call site,
$k$ is introduced existentially in the constraint, and left to be specified
later at the whims of the constraint solver. Lower values of $k$ lead to better
bounds, but must remain high enough to run all branches.

\subsection{Potential in Shared Values}

AARA usually stores potential within shared values as opposed to a centralized
token. We take a somewhat different approach to sharing: we want shared values
$!A(\Theta)$ to specify how much potential their instances commonly occupy, and
store those resources in the token. To do so, we define below a type
$!_{\phi}A(\Theta)$ which represent shared copies with potential. Building the
value requires a token with $\phi(\Theta)$ free resources and allocates them;
Extracting the underlying $!A(\Theta)$ from $!_{\phi}A(\Theta)$ frees
$\phi(\Theta)$ resources. This is definable without any primitive in the
\textbf{ILL}-machine, together with resource-aware contraction and weakening,
which fully reproduce AARA potential.
\[ !_{\phi} A(\Theta) = \closure \forall p,q.\; \ST(p,q +\phi(\Theta)) \lol \thunk \; \ST(p+\phi(n),q)\, \otimes\; !\, \thunk A(\Theta).\]

\section{Implementation for ML-like Languages}\label{sec:application}

We have implemented a
prototype\footnote{\url{https://gitlab.lip6.fr/suzanneh/autobill}} of the
\textbf{ILL}-machine, together with type inference and constraint generation
using the HM(X) technique~\cite{pottier2005}. HM(X) (\emph{Hindley-Milner
  extended with X}), is a generic constraint-based type inference procedure
extending Hindley-Milner with user-definable sorts, types and predicates. Our
extension of covers arbitrary first-order signatures for parameters, and
features a generic constraint simplifier. It then can export simplified
constraints to the \emph{Coq} theorem prover for verification by hand, or to the
\emph{MiniZinc}\footnote{\url{https://www.minizinc.org}} optimization suite to
for full cost inference with minized slack. This yields complexity bounds as a
multivariate polynomial. \diff{}{ Our heuristic for elaborating polynomial
  parameter expressions from a first-order constraint works as following:
\begin{enumerate}
  \item Take as input a first-order constraint over the integers. Its syntax is
        generated by $(\forall, \exists, \wedge, \Rightarrow, =, \le)$
  \item Skolemize all existential variables: $\exists y. C$ generates a fresh
        multivariate polynomial $p(\vec{x})$ over the variables $\vec{x}$ in
        scope, and reduces to $C[p(\vec{x})/y]$. Those polynomial $p$ are formal
        sums of monomials with coefficients $\vec{\alpha}$, which are all held
        in a global context for the polynomials.
  \item Assume all implications are of the form
        $(p(\vec{x}) = e[\vec{x}]) \Rightarrow C$, and substitute $p$ for $e$ in
        $C$
  \item Put the constraint under prenex form. We have arrived at a constraint
        $C = \forall \vec{x}. \bigwedge_{i}(e_{i}[\vec{x},\vec{\alpha}] = e'_{i}[\vec{x},\vec{\alpha}])$.
        Reinterpret the constraint as a system of polynomial equations with
        variables $\vec{x}$ and unknown coefficients $\vec{\alpha}$.
  \item Finally, instantiate and optimize $\vec{\alpha}$ under this final system of
        equations. The metric for the optimization is the sum of the leading
        non-zero coefficients of the complexity being computed.
\end{enumerate}}

Our preliminary experiments indicate that when manually annotating data\-types
definitions with their RAML parameterization, \diff{tight algorithmic
  complexities can be derived as in RAML for representative programs.}{the tight
  algorithmic complexities derivable for list iterators can be recovered in
  \textbf{BILL}. We are currently exploring ways to extend parameterization
  to tree-shaped datatypes, as well as the potential precision gains that can be
  obtained by parameterizing to Call-By-Push-Value evaluation
  contexts.}

The direct translation from ML-style languages to \textbf{ILL}-with-token is
factored in the implementation. User-facing languages are be translated to a
Call-By-Push-Value $\lambda$-calculus the canonical way, and then compiled into
\textbf{ILL} through a CPS-translation and explicit duplication of shared
variables. Later passes implement the Call-By-Push-Value effects described in
this paper, to-and-from \textbf{ILL}. High-level languages can be analyzed
more easily, as they only need to be translated to a Call-By-Push-Value
$\lambda$-calculus with credit/debit primitives.

\subsubsection{Limitations}

Higher-order functions are a pain-point for AARA analyses, as getting correct
bounds require lifting constraints out of high-order arguments. Depending on the
theory modeling parameters, this can be quite tricky. Our implementation
compares favorably to RAML in this regard, as it supports native constraints
annotations on high-order arguments. Fixpoints are also a thorny issue. This has
two mitigations: (1) require user-provided annotations for fixpoint invariants,
or (2) \diff{reduce usage to only tractable fixpoints, such as folds or
  traversal.}{reproduce RAML's constraints on iteration: (mutually) recursive
  functions which define folds and traversals through nested pattern-matching and
  accumulation.} In our experiments, we found our system to be amenable to a
\diff{mixed}{third} approach: general purpose iterators can be defined with
manual annotations using fixpoints, and then used without annotations.

\section{Conclusion}\label{sec:conclusion}

Extending the static type discipline of functional languages for resource
analysis is a tantalizing prospect. But understanding the operational properties
of programs means recovering a diverse swath of information like memory
aliasing, algorithmic invariants, and sharing of data outside their scope of
definition.

To create a fine-grained, generic, extendable base for AARA, we extended the
\textbf{ILL} Call-By-Push-Value calculus developed by Curien at
al.~\cite{munch-maccagnoni2017,curien2016} with fixpoints, polymorphism and
native first-order constraints. We combined this with a decomposition of
$\lambda$-calculi in a Call-By-Push-Value machine, which explicits control flow.
With this, expressions are a combination of closures requiring some properties
to hold on their inputs, and thunks which witness some properties of their
output. Instantiating our generic system to represent a finite amount of
resources within the program's state, well-typedness stipulates that this finite
amount is sufficient to cover all allocations and liberations. It recovers the
core of the AARA~\cite{hoffmann2022} method for resource analysis from first
principles, from a generic Call-By-Push-Value intermediate representation for
static analysis.

\subsubsection{Perspectives}
Our implementation covers the target machine, the constraint-aware type system,
and a heuristic to solve constraints over multivariate polynomial. Current work
focuses on implementing program-wide analysis \emph{à la} RAML. This requires
automatically annotating the constraints associated to each constructor in
datatype definitions to encode a particular flavor of AARA analyses. We also
aim to support shared regions, reusing the parameterized type system to generate
constraints on region lifetimes.

\bibliographystyle{splncs04}
\bibliography{LOPSTR23}

\begin{thebibliography}{10}
\providecommand{\url}[1]{\texttt{#1}}
\providecommand{\urlprefix}{URL }
\providecommand{\doi}[1]{https://doi.org/#1}

\bibitem{curien2016}
Curien, P.L., Fiore, M., {Munch-Maccagnoni}, G.: A theory of effects and
  resources: Adjunction models and polarised calculi. In: Proceedings of the
  {{ACM}} on {{Programming Languages}} ({{POPL}}) (Jan 2016).
  \doi{10.1145/2837614.2837652}

\bibitem{dallago2011}
Dal~Lago, U., Gaboardi, M.: Linear {{Dependent Types}} and {{Relative
  Completeness}}. In: 2011 {{IEEE}} 26th {{Annual Symposium}} on {{Logic}} in
  {{Computer Science}} (Jun 2011). \doi{10.1109/LICS.2011.22}

\bibitem{dallago2013}
{Dal lago}, U., Petit, B.: The {{Geometry}} of {{Types}}. ACM SIGPLAN Notices
  (POPL)  (Jan 2013). \doi{10.1145/2480359.2429090}

\bibitem{hoffmann2011}
Hoffmann, J.: Types with Potential: Polynomial Resource Bounds via Automatic
  Amortized Analysis. Ph.D. thesis, Ludwig-Maximilians-Universit\"at M\"unchen,
  {Berlin} (2011), \url{https://doi.org/10.5282/edoc.13955}

\bibitem{hoffmann2017}
Hoffmann, J., Das, A., Weng, S.C.: Towards automatic resource bound analysis
  for {{OCaml}}. In: Proceedings of the {{ACM}} on {{Programming Languages}}
  ({{POPL}}) (2017). \doi{10.1145/3009837.3009842}

\bibitem{hoffmann2022}
Hoffmann, J., Jost, S.: Two decades of automatic amortized resource analysis.
  Mathematical Structures in Computer Science pp. 1--31 (Mar 2022).
  \doi{10.1017/S0960129521000487}

\bibitem{kahn2020}
Kahn, D.M., Hoffmann, J.: Exponential automatic amortized resource analysis.
  In: International Conference on Foundations of Software Science and
  Computation Structures. pp. 359--380. {Springer, Cham} (2020)

\bibitem{levy2003}
Levy, P.B.: Call-{{By-Push-Value}}. {Springer Netherlands}, {Dordrecht} (2003).
  \doi{10.1007/978-94-007-0954-6}

\bibitem{lichtman2017}
Lichtman, B., Hoffmann, J.: Arrays and {{References}} in {{Resource Aware ML}}.
  In: Miller, D. (ed.) International {{Conference}} on {{Formal Structures}}
  for {{Computation}} and {{Deduction}} ({{FSCD}}) (2017).
  \doi{10.4230/LIPIcs.FSCD.2017.26}

\bibitem{lopez-garcia2018}
{Lopez-Garcia}, P., Darmawan, L., Klemen, M., Liqat, U., Bueno, F.,
  Hermenegildo, M.V.: Interval-based resource usage verification by translation
  into {{Horn}} clauses and an application to energy consumption. Theory and
  Practice of Logic Programming  \textbf{18}(2),  167--223 (Mar 2018).
  \doi{10.1017/S1471068418000042}

\bibitem{munch-maccagnoni2017}
{Munch-Maccagnoni}, G.: Note on curry's style for linear call-by-push-value.
  Tech. rep. (May 2017), \url{https://hal.inria.fr/hal-01528857}

\bibitem{niu2018}
Niu, Y., Hoffmann, J.: Automatic {{Space Bound Analysis}} for {{Functional
  Programs}} with {{Garbage Collection}}. In: {{LPAR-22}}. 22nd {{International
  Conference}} on {{Logic}} for {{Programming}}, {{Artificial Intelligence}}
  and {{Reasoning}}. pp. 543--521. \doi{10.29007/xkwx}

\bibitem{perez2020}
P{\'e}rez, V., Klemen, M., {L{\'o}pez-Garc{\'i}a}, P., Morales, J.F.,
  Hermenegildo, M.: Cost {{Analysis}} of {{Smart Contracts Via Parametric
  Resource Analysis}}. In: Pichardie, D., Sighireanu, M. (eds.) Static
  {{Analysis}}, vol. 12389, pp. 7--31. {Springer International Publishing},
  {Cham} (2020). \doi{10.1007/978-3-030-65474-0_2}

\bibitem{pottier2005}
Pottier, F., {Didier R\'emy}: The essence of {{ML}} type inference. In:
  {Pierce, Benjamin C} (ed.) Advanced {{Topics}} in {{Types}} and {{Programming
  Languages}}, pp. 389--489. {The MIT Press} (Jan 2005)

\bibitem{rajani2021}
Rajani, V., Gaboardi, M., Garg, D., Hoffmann, J.: A unifying type-theory for
  higher-order (amortized) cost analysis. Proceedings of the ACM on Programming
  Languages (POPL) (Jan 2021). \doi{10.1145/3434308}

\bibitem{tarjan1985}
Tarjan, R.E.: Amortized {{Computational Complexity}}. SIAM Journal on Algebraic
  Discrete Methods  (Apr 1985). \doi{10.1137/0606031}

\end{thebibliography}

\end{document}